# Phase-Gradient Huygens' Metasurface Coatings for Dynamic Beamforming in Linear Antennas

Stefano Vellucci, *Member, IEEE*, Michela Longhi, *Member, IEEE*, Alessio Monti, *Senior Member, IEEE*, Mirko Barbuto, *Senior Member, IEEE*, Alessandro Toscano, *Senior Member, IEEE*, and Filiberto Bilotti, *Fellow, IEEE*

*Abstract*— The beamforming capabilities of conformal cylindrical Huygens metasurface (HMS) coatings for linear antennas are assessed. It is shown that by engineering the phase-gradient profile of the HMS, the original omnidirectional radiation pattern of the linear antenna can be shaped to form multi- or single-beam configurations. A closed-form expression for the phase-insertion profile of the cylindrical coating required to achieve the desired radiation pattern profile is derived, and several full-wave numerical examples supporting our claims are reported. A configuration exploiting a realistic HMS layout is also discussed and it is shown that, by making the metasurface reconfigurable through the use of tunable lumped elements, the radiated beams can be dynamically steered in space. This new design methodology could find application in smart electromagnetic environment scenarios for dynamically rerouting the signal towards multiple users.

## I. INTRODUCTION

WIRELESS systems for next-generation communications (*e.g.*, *B5G, 5G+, 6G*) will face more stringent requirements compared to the current standard, relying on an increased number of connections, expanded user density, higher data transfer, lower latency, almost zero jitters, broadband connectivity, and massive reliability [1],[2]. Since these key performance indicators (*KPIs*) should be achieved while retaining low costs and minimum infrastructure complexity [3],[4], communication schemes exploiting carriers at sub-6 GHz and millimeter wave frequencies are expected to be the key pillar of future wireless services [5],[6]. Still, with the increase of the operative frequencies, the *KPIs* achievement is challenged by several detrimental factors due to the environment. Among them, path loss, blockage, and undesired scattering threaten the link budget in both outdoor and indoor scenarios. Unfortunately, attenuation experienced by the signals becomes severe at higher frequencies and countermeasures able to guarantee a stable and reliable quality of service are mandatory [7],[8]. Indeed, conventional solutions exploited in previous wireless communications generations, such as the network densification through an increase in the number of base-stations (*BTS*), or the boost of the antennas' *EIRP* (effective isotropic radiated power), are not preferred solutions due to the cost increase, massive power consumption, and alarming *EMF* (electromagnetic field) levels [9],[10]. A paradigm change, thus, is required through a modification of the communication scheme.

A promising solution to address these issues relies on the concept of *smart electromagnetic (EM) environment* [11]-[18]. In this framework, the environment is exploited as a new degree of freedom in the design of the communication scheme, rather than an impairment, and the propagation scenario plays a core role in mitigating losses, distortion and fading of the *EM* waves. This new paradigm is enabled by the development of different design strategies and innovative technologies, which introduce elements with "smartness" embedded at the physical layer, without requiring time-consuming and high-latency post-processing of network virtualized functions (*NTF*) [19].

Toward this end, a first successful example can be found in the so-called *smart skins* [14],[20],[21], which are static and low-cost metasurfaces that allow for a judicious manipulation of the reflected *EM* waves, even out of the classical Snell's laws, for covering spots and improve the non-line-of-sight performance. The feasibility of this solution is however paid with the absence of reconfigurable capabilities that limits their range of applications. To overcome this limitation, recently, extensive research efforts have been focused in the design of *reconfigurable-intelligent surfaces* (*RISs*) [12],[18],[22]-[26]. Such a technology exploits engineered metasurfaces covering buildings, walls, etc. that are made reconfigurable by means of electronic components. In this way, the incoming *EM* waves can be adaptively reflected towards anomalous angles. The tunability of the reflected wave's direction allows for easy beam management, making *RISs* a powerful tool to expand the

Manuscript received February 08 2023. This work has been developed in the frame of the activities of the Project MANTLES, funded by the Italian Ministry of University and Research under the PRIN 2017 Program (protocol number 2017BHFZKH). (Corresponding author: Stefano Vellucci).

S. Vellucci, M. Longhi, and M. Barbuto are with the Department of Engineering, Niccolò Cusano University, 00166, Rome, Italy (e-mail: stefano.vellucci@unicusano.it).

A. Monti, A. Toscano, and F. Bilotti are with the Department of Industrial, Electronic and Mechanical Engineering, ROMA TRE University, 00146 Rome, Italy.

channel capacity. Still, the presence of electronic components requires highly performing manufacturing, increased installation costs, and proper power management.

Finally, to further assist the smartness of the communication scheme, *smart repeaters* and *antennas* with dynamic amplification and forward control capability are introduced in the smart *EM* environment scenario [27]. These higher level smart nodes are expected to further enhance the coverage capability of the network, boosting the signal strength, and introducing reconfigurable re-transmission capabilities. Compared to current technologies, where the repeaters are characterized by an omni or fixed directionality, *i.e.*, have transmission and reception characteristics fixed over time, next-generation smart antennas will require multi-beam capability and adaptability to maximize the data link performance with the final user [19], in both outdoor and indoor scenarios. Thus, antenna solutions introducing reconfigurability and shaping of the radiation pattern, scattering manipulation, and/or tuning of the frequency of operation may be profitably considered.

In this general context, a class of antennas covered by conformal and flexible metasurface sheets enabling advanced control over these characteristics has been proposed in the last decade [28],[29]. In these cases, the metasurface is used as a shell wrapped around a linear antenna introducing advanced wave-manipulation functionalities, albeit allowing for proper reception/transmission of the incoming signal (*i.e.*, not isolating the antenna from the surrounding environment). For instance, one of the first devices exploiting this design strategy relies on the use of coating metasurfaces for cloaking functionalities, enabling the design of extremely compact telecommunication platforms equipped with multiple antenna systems confined in a reduced space [30]-[34]. Similar principles have been also applied for broadening or tuning the antenna's resonance frequency by a judicious variation of the wrapping metasurface response [35]-[37]. Finally, by utilizing non-linear electronic devices loading the metasurface, antennas and array systems exhibiting different radiation characteristics depending on the power level or even the signal waveform have been discussed [38]-[42].

In this contribution, we aim to further expand the potentialities of metasurface shells coating wired antennas and being able to manipulate their scattering, frequency, or radiative characteristics. In particular, the design of a linear antenna wrapped by a cylindrical Huygens metasurface (*HMS*) is proposed. Thanks to the engineering of the phase-insertion introduced by the phase-gradient discontinuity, the radiation pattern of the antenna is shaped at will on the azimuthal plane. By tuning the surface impedance values of the *HMS*, *i.e.*, by modifying the phase-gradient profile, beamforming capability is thus introduced. A realistic *HMS* loaded with varactor diodes is then designed, enabling dynamic control over the radiation beam characteristics.

The paper is organized as follows: in Section II, the theory and working principle for achieving beam manipulation are presented, and the analytical formulation for evaluating the phase-insertion profile to be introduced by the gradient *HMS* is derived. The strategy for discretizing the metasurface depending on the desired radiation pattern shape is also discussed, as well as a complete design workflow for the unit-cells composing the *HMS* is introduced. In Section III, these design principles are applied to several numerical examples exploiting ideal *HMS* cells, showing potentialities and limitations of the proposed approach in obtaining single- or multi-beam configurations. Finally, in Section IV, we report the design of a realistic configuration made of a reconfigurable *HMS* loaded with varactor diodes enabling unprecedented dynamic beam-forming capabilities for wire antenna systems.

## II. Design Principles of the Cylindrical Huygens Metasurface Coating

In this Section, we introduce the use of a cylindrical *HMS* coating to manipulate the radiation pattern of a linear antenna. Albeit their simplicity, wired antennas such as dipoles, monopoles, strips, etc. are widely used in many applications (e.g., mobile communications, networking, sensing, IoT, etc.) because to their straightforward implementation, ease of fabrication, and low cost [43]. Nevertheless, the low directivity and omnidirectionality of the radiation pattern on the *H*-plane (*i.e.*, on the plane orthogonal to the wire direction) severely limits their range of applications. Hence, augmented linear antennas overcoming this limitation can be particularly appealing to antenna engineers, limiting the use of antenna arrays.

In this framework, recently, *HMSs* have shown remarkable possibilities in tailoring the *EM* wavefront. *HMSs* consist of unit-cells that guarantee the full transmission of the impinging field and full control over its phase through a proper combination of the electric and magnetic dipole moments [44]-[46]. The possibility to excite both the dipole moments and properly balance them distinguishes *HMSs* from conventional metasurfaces, where usually the control of either the electric or magnetic responses is not simultaneous [47]. Indeed, in *HMSs* the possibility of properly tailoring the electric and magnetic polarizabilities (or impedances) allows for a peculiar unidirectional radiation pattern. Intriguingly, the name "*Huygens*" stems from the perpendicular colocation of the electric and dipole moments, which allow to excite currents onto the discontinuity acting as secondary unidirectional sources, as envisioned in the Huygens principle [48].

Among the different possibilities enabled by *HMSs*, the design of phase-gradient metasurfaces for refracting in unconventional ways the incident field is one of the most discussed [49],[50]. In the antenna framework, these platforms have been especially used in phased array scenarios to introduce abrupt discontinuities in the phase of the propagating wave and attain interesting focusing or steering functionalities [51]-[55].

However, the steering and beamforming capabilities of phase-gradient *HMSs* can be extended also to the case of a single radiating source, as recently envisioned in our conference contribution [56]. In Fig. 1, the reference geometry considered here is reported. A conventional half-wavelength dipole antenna is surrounded by a conformal cylindrical *HMS* of radius $a_c$, which is designed to introduce a phase-gradient on the *azimuth* direction. Hence, the *HMS* consists of an anisotropic transition sheet with inhomogeneous unit-cell along the azimuth that aims at manipulating the original omnidirectional pattern of the antenna on the *H*-plane. Namely, a distributed phase-insertion on the *xoy*-plane is introduced, leading to a focusing effect and beamforming capabilities on the azimuth.

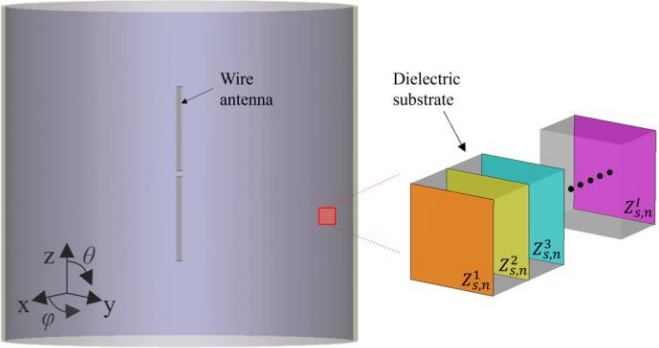

Fig. 1: Reference geometry considered in the paper consisting of a cylindrical phase-gradient *HMS* coating a linear dipole antenna. Through the engineering of the phase-gradient introduced by the *HMS* beam-steering capabilities are introduced. In the inset, a sketch of the unit-cell configuration used to implement the *HMS* is reported.

To better explain this point, a half vertical cross-section of the problem under consideration is reported in Fig. 2. According to the antenna theory [43], on the azimuthal-plane the phase front of the field radiated by the dipole is (*quasi-*) cylindrical, with a constant value of the phase along the cylindrical profile of the *HMS* coat. Therefore, at the *HMS* discontinuity, we have a stable value of the phase of the incoming wave point by point. To achieve the beamforming functionality, the *HMS* should tailor its phase-insertion by transforming the almost cylindrical wavefront on the horizontal plane into almost planar wavefronts. From Fig. 2, assuming a ray approach and radial propagation, it can be inferred that the *HMS* should compensate for the different electrical paths of the rays travelling from the omnidirectional source placed at the center. The engineered phase-insertion along the curvilinear profile that leads to the plane wave propagation can be calculated by using elementary trigonometric relations. In particular, the phase-gradient of the *HMS* assumes the following expression:

$$\Phi(\varphi) = \Phi_0 + k_0 d(\varphi) = \Phi_0 + k_0 a_c [1 - \cos^{-1} \varphi] \quad (1)$$

where $\varphi$ is the geometrical coordinate along the cylindrical profile, $\Phi_0$ is an arbitrary reference phase value, $k_0 = 2\pi/\lambda_0$ is the free-space wavenumber, and $d(\varphi)$ is the differential electrical path of the ray. It is worth noting that, to introduce a focusing effect, the phase-insertion $\Phi(\varphi)$ should be an even function. In other terms, the *HMS* introduces the same phase-insertion for positive and negative values of $\varphi$.

Arguably, the number of radiation beams generated exploiting this approach is regulated by the spatial periodicity of the phase-insertion function $\Phi(\varphi)$, which defines the number of sectors into which the *HMS* is divided, and where a periodicity of $2\pi$ corresponds to the full spatial coverage of the circular profile. Due to the symmetry of the problem and the even nature of $\Phi(\varphi)$, the largest periodicity value of the phase-insertion function is $\pi$. In this case, the same phase-gradient profile is distributed along a semi-circle (as in the scenario in Fig. 2), leading to a two sectors configuration and two main symmetrical radiating beams pointing at opposite directions. Indeed, this point and full-wave numerical results supporting our claim will be better discussed in the next section.

Once the profile of the continuous phase function $\Phi(\varphi)$ is derived, for practical implementation it needs to be discretized in a finite number of samples. Each discrete sample of the function corresponds to a unit-cell of the phase-gradient *HMS* and we assume a plane-wave impinging onto each cell. Thus, once the number (*i.e.* $2N - 1$) of unit-cells implementing one period of $\Phi(\varphi)$ is fixed, the required phase-insertion $\Phi(\varphi_n)$ exhibited by the *n*-th unit cell can be derived using (1) (with $n = 1, 2, \ldots N$).

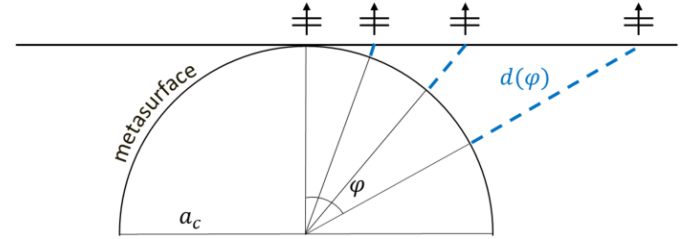

Fig. 2: Schematic representation on a 2D plane depicting the working principle of the cylindrical *HMS*. Radial propagation from the central source element is assumed. At the cylindrical discontinuity, the *HMS* introduces a phase-insertion able compensating for the different electrical paths of the rays $d(\varphi)$ (*blue dashed lines*) when focused to propagate towards the same direction. The original cylindrical phase-fronts are thus transformed into plane waves. $\varphi$ represents the angular position of the radiated ray.

For the design of the cells exhibiting the required $\Phi(\varphi_n)$, the configuration reported in the inset of Fig. 1 can be considered. Each cell is made of stacked ideal impedance sheets, characterized by a surface impedance value $Z_{s,n}^i$ with $i = 1, 2, \ldots I$, being $I$ equal to the number of stacked layers. The impedance sheets are separated by the same dielectric substrate of thickness *t* and relative permittivity $\varepsilon_r$, and the $Z_{s,n}^i$ is assumed to be purely reactive (*i.e.*, $Z_{s,n}^i = jX_{s,n}^i$ with *j* being the imaginary unit). This configuration enables the excitation of both the electric and magnetic dipole moments ensuring, ideally, high transmission values and a phase coverage that strongly depends on the dielectric thickness [57]. It is worth mentioning that, albeit two impedance sheets are generally enough to design an *HMS* cell [44], more layers are here considered to increase the phase coverage and the design degrees of freedom.

This stacked configuration can be modelled through a cascade of transmission matrices corresponding to the ABCD matrices of the different material layers and of the transition sheets [59]. Thus, the overall ABCD matrix of the *n*-th unit-cell can be written as:

$$\begin{pmatrix} A_n & B_n \\ C_n & D_n \end{pmatrix} = Y_{s,n}^1 \cdot d \cdot Y_{s,n}^2 \cdot d \ldots Y_{s,n}^I \quad (2)$$

where $Y_{s,n}^i$ and $d$ are the ABCD matrices of the admittance sheets and the dielectric substrates, respectively. For normal incidence, the latter reduces to [60][59]:

$$d = \begin{pmatrix} \cos(k_0 d\sqrt{\varepsilon_r}) & j\dfrac{\eta_0 \sin(k_0 d\sqrt{\varepsilon_r})}{\sqrt{\varepsilon_r}} \\ j\dfrac{\eta_0 \sin(k_0 d\sqrt{\varepsilon_r})}{\sqrt{\varepsilon_r}} & \cos(k_0 d\sqrt{\varepsilon_r}) \end{pmatrix} \quad (3)$$

with $\eta_0$ being the free-space wave impedance. If we also assume that the admittance sheets exhibit only an electric response and are isotropic on the plane of the interface, the ABCD matrix of a single admittance layer can be written as:

$$Y_{s,n}^i = \begin{pmatrix} 1 & 0 \\ 1/jX_{s,n}^i & 1 \end{pmatrix} \quad (4)$$

Finally, from the ABCD matrix (2), by using (3) and (4), the ABCD matrix of the single cell can be related to its transmission and reflection coefficients through the relations [59]:

$$S_{2,1}^n(Y_{s,n}^i, d) = \dfrac{2}{A_n + B_n/\eta_0 + \eta_0 C_n + D_n}$$
$$S_{1,1}^n(Y_{s,n}^i, d) = \dfrac{A_n + B_n/\eta_0 - \eta_0 C_n - D_n}{A_n + B_n/\eta_0 + \eta_0 C_n + D_n} \quad (5)$$

The transmission (reflection) coefficient values, in both amplitude and phase, of the single *HMS* unit-cell are thus controlled by the dielectric material properties and the surface impedances of the ideal impedance sheets. It is worth noting that (5) uses the canonical scattering parameters transformation, assuming a moderate value of $\varphi$. Conversely, the generalized scattering parameters [61] should be eventually used to properly satisfy the local power conservation requirement ensuring the matching of the different wave impedances at the cell sides [58].

Through the above formalism, we can design the $2N - 1$ discrete cells that guarantee the desired phase-insertion profile resulting from (1). Indeed, once the dielectric characteristics and the number of layers $I$ are fixed, (5) can be used to find the proper combinations of the surface impedance values $Z_{s,n}^1$, $Z_{s,n}^i, \dots, Z_{s,n}^I$ satisfying the following conditions:

$$arg(S_{2,1}^n) = \Phi(\varphi_n), \quad |S_{2,1}^n| = 1 \quad (6)$$

From a theoretical point of view, the availability of an analytical model allows the synthesis of the surface impedance values of the *n*-th unit-cell composing the *HMS*. However, the complexity of the analytical formulation and the underdetermined nature of the resulting algebraic system make this approach not feasible. Therefore, as described in [60], for the design of the *n*-th unit-cell satisfying (6), the cascade matrix model is used to numerically compute a database of transmission coefficient values of a single stacked *HMS* unit cell for different combinations of $Z_{s,n}^1$, $Z_{s,n}^i, \dots, Z_{s,n}^I$. Finally, among all the solutions returned, only the ones satisfying (6) are considered.

## III. NUMERICAL EXAMPLES OF BEAMFORMING HUYGENS METASURFACE COATINGS

To prove the validity and the potentiality of this approach, here, we apply the proposed design workflow to several study cases. In particular, the synthesis of different cylindrical *HMSs* aiming at tuning the *EM* radiation characteristics of a coated antenna are reported. The formalism previously presented is discussed when considering ideal lossless impedance transition sheets (*i.e.*, $Z_{s,n}^i = jX_{s,n}^i$ is assumed), and the numerical results for the different configurations are reported.

### A. Four-sector antenna

As a first example, we detail the synthesis of a cylindrical *HMS* able to focus the field radiated by a linear antenna source towards four symmetric directions. The *HMS* is divided into four focusing sections characterized by the same phase-gradient profile, as shown in Fig. 3. Thus, the spatial periodicity of $\Phi(\varphi)$ is $\pi/2$. As a reference, the feeding antenna is designed to work at the central frequency $f_0 = 2.5$ GHz (albeit the design can be easily scaled at other frequencies), while the *HMS* radius is $a_c = \lambda_0/2$. This value is chosen as a trade-off between the space occupancy of the coating and the minimum distance allowing the *HMS* to work properly. Further reducing the coating aspect ratio would put the *HMS* in the very near-field of the antenna, massively compromising the *HMS* response. Indeed, when placing the *HMS* in the reactive region of the antenna (*i.e.,* before the Fresnel region [43]) colocation of the electric and magnetic dipole moments is quite difficult. Moreover, as can be inferred from Fig. 2, extremely small unit-cells would be required to sample uniformly the coating metasurface and guarantee a smooth discretization of $\Phi(\varphi)$, due to the reduced *HMS* area.

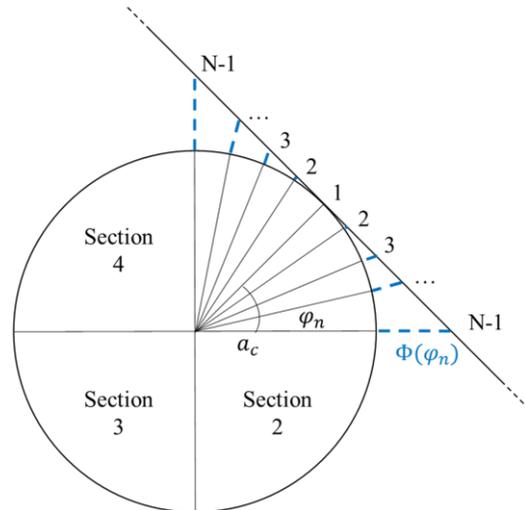

Fig. 3: 2D geometrical sketch the division in sections and sectors of the cylindrical *HMS* returning a four-beam symmetrical radiation pattern. Each section is discretized through *N* unit-cells introducing a phase-shift $\Phi(\varphi_n)$. $\varphi_n$ represents the angular position of the *n*-th unit-cell *w.r.t.* the centre of the section.

For the implementation of the metasurface, each section is also discretized in $2N - 1 = 15$ unit-cells providing the desired phase-insertion, *i.e.*, ensuring the required phase shift for each of the $n$-th cells. The latter is evaluated by introducing the referenced $f_0$ and $a_c$ values in (1), leading to $\varphi_n = n\varphi = n6°$ and, thus, with the discretized phase-insertion function $\Phi(\varphi_n)$ introducing a phase coverage $\Delta\Phi(\varphi_n) = |\Phi(\varphi_8) - \Phi(\varphi_1)| = 62°$, as resumed in Fig. 4.

Finally, exploiting the matrix representation and formalism reported in (5), it is found that this phase range is easily covered by unit-cells consisting of two dielectric substrates ($t = \lambda_0/60$, $\varepsilon_r = 10$) sandwiched between three purely reactive transition sheets. In particular, the combinations of surface reactances $X_{s,n}^1$, $X_{s,n}^2$, $X_{s,n}^3$ satisfying the first equation in (6) with $|S_{2,1}^n| \geq 0.9$ are in the range [-990, +100] $\Omega$/sq. The different values of the surface impedances used to implement the 15 unit-cells and the corresponding phase-insertions are reported in Fig. 4.

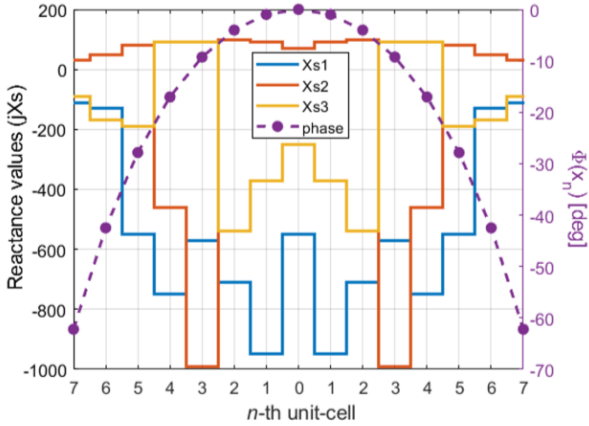

Fig. 4: (*Continuous lines*) Reactance values (imaginary parts) of the first ($X_s^1$), second ($X_s^2$), and third ($X_s^3$) impedance layers of the metasurface discretized in 15 unit-cells for the four-beam *HMS* case. (*Dashed line*) Phase-insertion returned by the 15 unit-cells. The $X_s$ measurement unit is $\Omega$/sq.

The analytically designed unit-cells are implemented in a full-wave *EM* simulation tool as subwavelength constituents of the cylindrical *HMS* coating. Since the *HMS* aims at introducing a phase-gradient only on the *H*-plane, the individual *HMS* unit-cells are invariant along the *E*-plane, as can be observed in the sketch reported in the inset of Fig. 5. In other terms, a vertically homogeneous distribution of square cells is assumed along the axis of the cylindrical coating.

Finally, in Fig. 5 the 2D gain behavior of the designed phase-gradient *HMS* is reported and compared to the uncoated antenna scenario. The gain polar plots are evaluated on both *H*- and *E*-planes, with the latter considered along the plane of maximum directivity (*i.e.*, for $Phi = 45°$). Thanks to the focusing effect coming from the four metasurface sections, four main radiating beams are observed, and the original constant gain value of the uncoated antenna increases to 4.98 dBi. At the same time, due to the absence of a phase-gradient along the *z*-axis, the original *inverted-8* shape of the radiation pattern is unaltered on the *E*-plane, though with a different absolute value, due to the power-conservation principle.

It is worth noting that by reconfiguring the sheet reactance values of the unit cells, the beams rotate and a dynamic coverage on the horizontal plane is enabled. Arguably, a similar behavior can be assessed either using four independent radiating elements (see as an example [62]) or by employing electronically steerable parasitic array radiator antennas (ESPAR) [63]. Still, compared to the former solution, here, the rotation of the beams is obtained through a single radiator coated by the proposed *HMS*. Whilst in the latter case, the use of parasitic radiators increasing the antenna space occupancy and loaded with high-value complex reactive load controls can be required. Moreover, the limited number of passive elements that can be used can further limit the scanning spatial resolution.

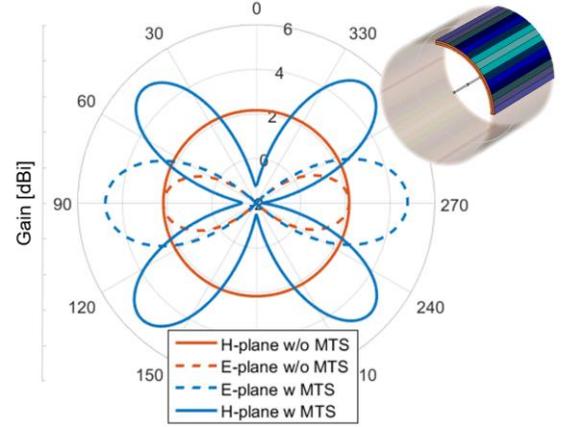

Fig. 5: Gain polar plots of the four-beam *HMS* coated antenna and of the bare dipole (*blue and red lines*, respectively) on the *H*-plane (*continuous line*) and *E*-plane (*dashed line*). In the inset, the sketch of the coated antenna highlighting a single *HMS* sector is reported. The different colours of the unit cells pictorially represent the metasurface phase-gradient in the azimuthal direction.

### B. Two-sector antenna

To demonstrate the versatility of our approach in achieving beamforming functionality, we also report another design case considering a dual-beam *HMS* coated linear antenna. The design procedure follows the one described in the previous case. However, here, the spatial periodicity of the phase-insertion function $\Phi(\varphi)$ is set to $\pi$. Namely, the cylindrical metasurface is divided into two equal surfaces introducing the same phase-gradient distribution, as pictorially represented in the inset of Fig. 6.

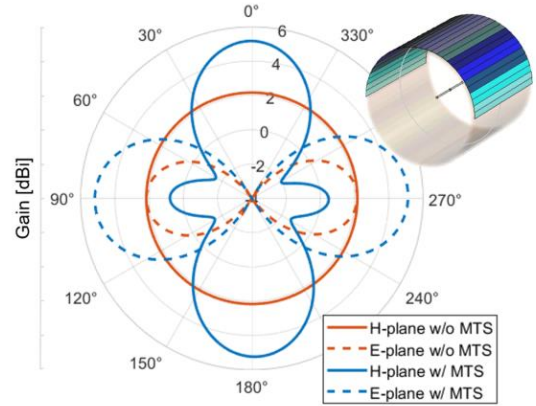

Fig. 6: Gain polar plots of the two-beam *HMS* coated antenna and of the bare dipole (*blue and red lines*, respectively) on the *H*-plane (*continuous line*) and *E*-plane (*dashed line*). In the inset, the sketch of the coated antenna highlighting a single *HMS* sector is reported. The different colours of the unit cells pictorially represent the metasurface phase-gradient in the azimuthal direction.

To keep the same unit-cell dimension of the previous scenario, guaranteeing proper homogenization of the metasurface, the *HMS* is discretized in $2N - 1 = 30$ unit-cells. The phase-insertion is still evaluated through (1), returning a discretized phase-insertion function $\Phi(\varphi_n)$, requiring a phase coverage in the range $\Delta\Phi(\varphi_n) = |\Phi(\varphi_{15}) - \Phi(\varphi_1)| = -740°$. Thus, the unit-cell configuration should be able to cover a full $2\pi$ phase range. Namely, the same three-layered structure previously considered is used, and the $X_{s,n}^1$, $X_{s,n}^2$, $X_{s,n}^3$ are synthetized selecting, from the database of combinations compliant with the first equation in (6), the ones still providing $|S_{2,1}^n| \geq 0.9$. Here, the surface reactance values composing the three layers of the metasurface and returning the desired phase coverage are in the range [-940, +90] $\Omega$/sq.

The full-wave numerical results of the designed structure are reported in Fig. 6. The polar plot gain shows two main lobes pointing in opposite directions on the *H*-plane, with a maximum gain of 5.17 dBi. As in the previous example, the shape of the radiation pattern remains almost unaltered on the *E*-plane.

### C. Single-sector scanning antenna

Though in the previous examples we focused on devices leading to multi-beam configurations, the coating metasurface can be also designed to generate a single radiating beam that can be used for scanning on the *H*-plane.

A sketch of this antenna is depicted in the inset of Fig. 7. The cylindrical coating is divided into two symmetric sectors, as in the previous example, but one sector is characterized by perfectly reflective unit-cells (the blue sector in Fig. 7), whilst the other one is designed following the procedure used so far for the design of a *HMS* (the orange sector in Fig. 7). The latter section relies on the use of a full-transmitting semi-cylindrical *HMS*, whilst the former is constituted by *n*-th reflective unit-cells with the same values of $X_s^1$, $X_s^2$, $X_s^3$. In this case, the surface reactances are found exploiting the reflection coefficient formula in (5) and imposing the condition:

$$arg(S_{1,1}^n) = \Phi' = 2(\pi - k_0 a_c), \qquad |S_{1,1}^n| \cong 1 \qquad (7)$$

Equation (7) ensures constructive interference between the field reflected by the reflecting sector and the one transmitted through the *HMS* sector. When considering $a_c = \lambda_0/2$, eq. (7) returns $arg(S_{1,1}^n) = 0$ and the surface reactance values satisfying this condition are $X_s^1 = -72$, $X_s^2 = 72$, $X_s^3 = -6$ $\Omega$/sq.

It is worth noticing that the condition $arg(S_{1,1}^n) = 0$ differs from the one of a conventional reflector made of a perfect electric conductor (PEC) where the distance between the antenna and the reflector should be $d = \lambda_0/4$ [43] (which in (7) leads to the condition $arg(S_{1,1}^n) = \pi$). Importantly, eq. (7) is not restricted to a specific value of $a_c$ but allows designing a cylindrical reflective panel for any distance value, increasing the degree of freedom compared to conventional metallic reflective panels.

Finally, in Fig. 7, the gain performance of the antenna coated by the cylindrical metasurface is reported. As can be appreciated from the polar plot, a directive beam appears on the *H*-plane with a maximum gain of 7.33 dBi. Remarkably, by making the metasurface unit-cell reconfigurable, it can be envisioned a system where the radiation characteristics of the antenna can be switched from the ones of an omnidirectional radiator to the ones of a directive transmitting/receiving antenna system able to scan the horizontal plane. In the next Section, a reconfigurable system is designed to address this vision.

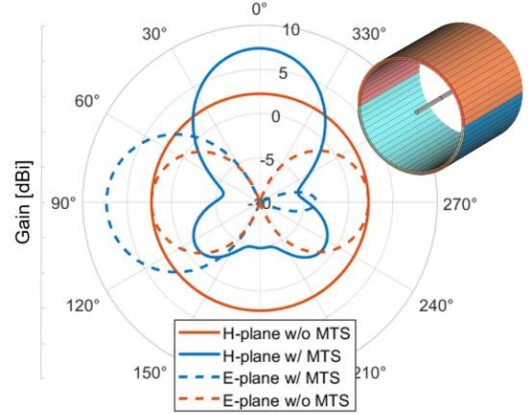

Fig. 7: Gain polar plots of the single-beam *HMS* coated antenna and of the bare dipole (*blue* and *red lines*, respectively) on the *H*-plane (*continuous line*) and E-plane (*dashed line*). In the inset, the sketch of the coated antenna highlighting the reflective and transmitting sectors are reported (*blue* and *orange* semi-cylindrical surfaces, respectively).

To conclude this Section, we report some insightful remarks on the relation between the directivity of the beamformed radiation pattern and the *HMS* aspect ratio. In Fig. 8, the results for three different configurations of the four-beam *HMS* coated linear antenna are reported. The metasurface coatings differ in the value of the parameter $a_c$, *i.e.*, for the dimension of the *HMS* radius. In particular, the cases for $a_c = \lambda_0/2, \lambda_0, 2\lambda_0$ are reported. The coating layers have been designed following the conventional procedure and evaluating (1) considering the three different values of $a_c$. Moreover, the same unit-cell dimension has been considered for the different configurations. Thus, due to the increase of the *HMS* area when increasing its radius, each focusing sector has been implemented with $2N - 1 = 15, 30$ and $60$ cells, respectively.

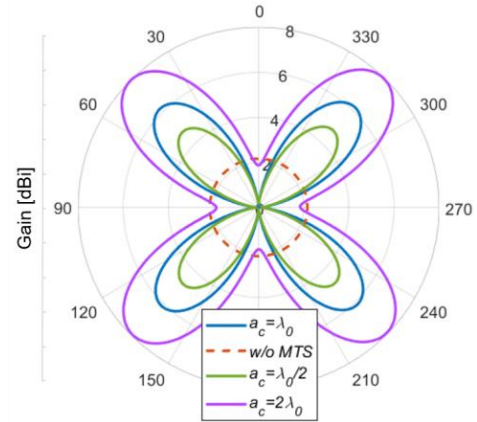

Fig. 8: Gain polar plots evaluated on the *H*-plane for different four-beam *HMS* coated antennas when the *HMS* radius is $a_c = \lambda_0/2, \lambda_0, 2\lambda_0$ (*continuous green, blue* and *purple lines*, respectively). Bare antenna (*red dashed line*).

From Fig. 8, a massive increase in the maximum directivity value can be observed. Arguably, the increase in the directivity comes from the extended equivalent radiating area of the coating metasurface, demonstrating that a trade-off between the maximum gain and the *HMS* aspect ratio is always needed.

## IV. REALISTIC CONFIGURATION OF A DYNAMIC BEAMFORMING LINEAR ANTENNA

We have shown that, by properly tuning the phase-insertion profile of a cylindrical *HMS* along the azimuthal plane, the radiation pattern of a linear antenna can be shaped in different ways. Since the $\Phi(\varphi)$ profile is controlled by modifying the responses of the individual unit-cells, *i.e.*, it can be tuned by controlling the surface impedance values of the cells, here, we explore the possibility to add reconfigurability to a preselected beamforming system. In particular, we focus our attention to the realistic configuration of the multi-beam device reported in Fig. 5, with the aim of implementing an antenna radiation pattern that can be dynamically switched between different configurations. For this purpose, the design of realistic and feasible unit-cells is considered, as well as the use of varactor diodes to tune the phase-insertion provided by the *HMS* cells.

One of the key aspects when dealing with metasurfaces equipped with varactor diodes is the minimization of the number of electronic elements to reduce the losses and the overall cost of the structure. At the scope, hereinafter, we consider unit-cells implemented through two-impedance sheets sandwiching a dielectric spacer, where the top layer is loaded with a varactor to tune the phase-insertion introduced by the cell. In other terms, the cell consists of a top dynamic layer characterized by a variable $X_s^1$, modified thanks to the varactor loading, and a bottom static layer with a fixed value of the $X_s^2$. This structure ensures the minimization of the number of electronic elements required to modulate the $\Phi(\varphi)$. Moreover, to further minimize the number of varactors, the *HMS* is implemented through a coarse discretization of the unit-cells. Namely, $2N - 1 = 5$ cells are used for one metasurface sector (*i.e.*, for each $\Phi(\varphi)$ period), leading to square unit cells having a length $l = \lambda_0/5$. It is worth mentioning that, though the discretization is rather course, the effectiveness of this strategy has been recently demonstrated in [66].

As a first step of the design, the characteristics of the dielectric substrate are defined. In particular, the commercial dielectric Rogers RO3006 ($\varepsilon_r = 6.15$, $\tan\delta = 0.002$) is selected for its flexibility towards bending, which makes it particularly suited for curved surfaces. To determine the value of the substrate thickness *t*, the $arg(S_{2,1}^n)$ coming from (5) for $\varepsilon_r = 6.15$ and $f_0 = 2.5$ GHz is evaluated when varying the parameters $t$, $X_s^1$ and $X_s^2$. Since to properly design a two-layer *HMS* unit-cell both inductive ($X_s \geq 0$) and capacitive ($X_s \leq 0$) sheets are required, the $X_s^1$ and $X_s^2$ are considered in the realistic ranges of [-1000, 0] and [0, +300] $\Omega$/sq, respectively, whilst $t$ varies according to the values available from the RO3006 datasheet.

In Fig. 9, the results filtered out for $|S_{2,1}^n| \geq 0.9$ are reported. For $t \geq 1.5$ mm a smooth phase variation within the range required for the four-beam *HMS* scenario (*i.e.*, $\Delta\Phi(\varphi_n) = 62°$) is guaranteed. Thus, $t = 1.5$ mm is selected as the thickness for the substrate. We remark that, although larger values of *t* guarantee even smoother coverage, a further increase of the dielectric thickness would compromise the flexibility of the substrate.

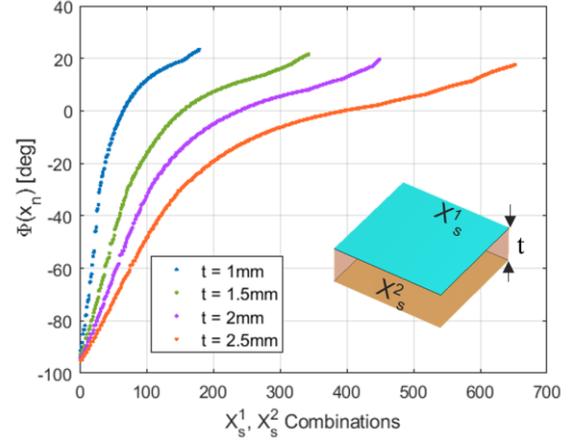

Fig. 9: Phase coverage $\Phi(\varphi_n)$ returned by (5) for a unit-cell made of a dielectric substrate with $\varepsilon_r = 6.15$ sandwiched between two impedance layers (in the inset) when considering different combinations of the surface reactances $X_s^1$ and $X_s^2$, and when varying the thickness $t$ of the spacer. The ranges considered for $X_s^1$ and $X_s^2$ are [-1000, 0] and [0, +300] $\Omega$/sq, respectively.

Once the characteristics of the substrate have been defined, the optimum value for the static surface impedance $X_s^2$ of the bottom layer is fixed. For this purpose, eq. (5) is once again evaluated but for $t = 1.5$ mm. The behavior of $arg(S_{2,1}^n)$ for fixed values of the $X_s^2$ and varying $X_s^1$ is reported in Fig. 10. As can be seen, the phase coverage massively depends on the static value of $X_s^2$. In particular, the widest phase coverage is achieved for $X_s^2 = 225$ $\Omega$/sq. This value guarantees, in principle, the full coverage of the phase-insertion profile in Fig. 4 when varying $X_s^1$ within realistic values. We stress that further increasing the value of $X_s^2$ would make the synthesis of a realistic inductive layer more difficult. Arguably, a high value of the $X_s^2$ requires the synthesis of an impedance layer characterized by an extremely high inductive response, which can be quite cumbersome to achieve when considering feasible layouts. Hence, the value of the $X_s^2$ is fixed to 225 $\Omega$/sq.

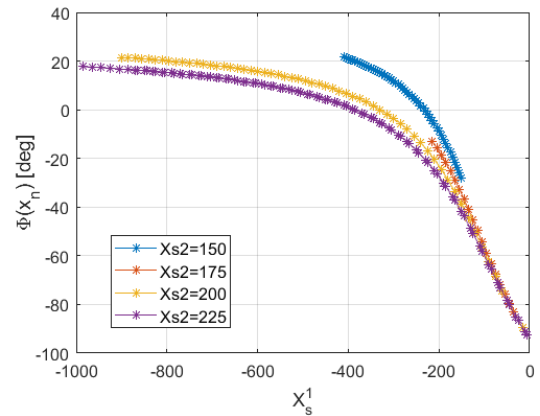

Fig. 10: Phase coverage $\Phi(\varphi_n)$ returned by (5) for a two-layer unit-cell separated by a dielectric substrate with $\varepsilon_r = 6.15$ and $t = 1.5$ mm, when $X_s^2$ is fixed and $X_s^1$ varies. The $X_s$ measurement unit is $\Omega$/sq.

To practically implement the inductive layer returning an $X_s^2 = 225$ Ω/sq, a simple and feasible layout consisting of a single metallic strip etched on the bottom face of the dielectric spacer is used, as shown in the inset of Fig. 11. As known [64],[65], this design returns a unit-cell exhibiting an inductive response when excited by a TE polarized wave ($E_z \neq 0, E_x = E_y = 0$). Exploiting the analytical formulas in [64] and considering the presence of the dielectric substrate and the unit-cell dimension, it is found that $X_s^2 = 225$ Ω/sq is achieved when the width of the metallic strip is $w_L = \lambda_0/160$.

Conversely, for the design of the capacitive layer, a metallic strip with a gap in the middle is considered. The gap is then loaded with a varactor to tune the surface impedance value, as depicted in the inset of Fig. 11. This configuration can be modelled as the parallel combination between the load capacitance and the equivalent capacitance introduced by the patterned surface $C_{mts}$. Namely, in the inset of Fig. 11, the circuit schematic of the unit-cell top layer is exemplified (neglecting the internal diode resistance $R_s$ for convenience). Here, $C_j$ represents the junction capacitance of the varactor, while $C_{pkg}$ and $L_{pkg}$ represent the parasitic reactive effects due to the packaging. The required junction capacitance $C_{j,n}$ to get the needed value of the $n$-th unit-cell $X_{s,n}^1$ can be thus evaluated as:

$$C_{j,n} = \frac{C_{mts} + C_{pkg} - C_n}{1 + (C_n - C_{mts} - C_{pkg})\omega^2 L_{pkg}} \quad (8)$$

where $C_n = -1/\omega X_{s,n}^1$. By defining the desired range of variation for $X_s^1$ and by fixing the maximum and minimum values of the $C_{j,n}$, the final dimensions of the capacitive strip can be derived. In particular, from Fig. 10, it can be inferred that to cover the required phase-insertion $\Delta\Phi(\varphi_n) = 62°$, $X_{s,n}^1$ should vary between $-1000 \leq X_{s,n}^1 \leq -100$. For the variation range of $C_{j,n}$, realistic values for the commercial varactor diode MGV125-08 from MACOM (with $L_{pkg} = 0.4$ nH, $C_{pkg} = 80$ fF) are considered. Namely, the varactor capacitance varies from about 0.6 to 0.05 pF when its reverse voltage changes from 2 to 20 V.

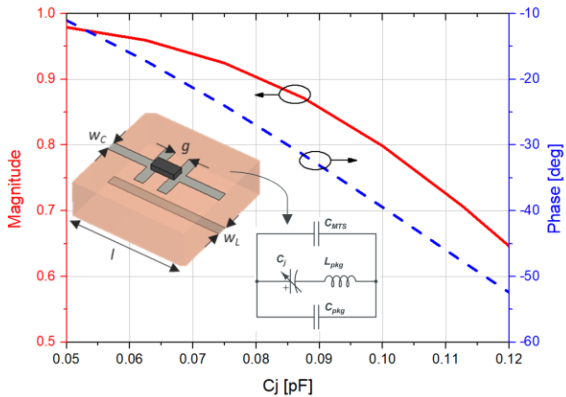

Fig. 11: Transmission coefficient values ($S_{21}$) in amplitude and phase of the reconfigurable *HMS* unit-cell when varying the junction capacitance ($C_j$) of the varactor. In the insets, a sketch of the varactor-loaded unit-cell and the equivalent circuit schematic of the top layer, are reported.

Hence, the final dimensions of the capacitive top strip are $w_C = \lambda_0/92$ and $g = \lambda_0/40$, and the transmission coefficient behavior for the unit-cell when excited by *Floquet* modes under TE illumination is reported in Fig. 11, in both amplitude and phase. As can be appreciated, when varying $C_j$ within [0.05 – 0.13 pF] the desired phase range is covered. However, due to the realistic configuration of the layout, the amplitude of the transmission coefficient degrades more significantly compared to the ideal setup, especially for large phase shift values. Anyway, we remark here that only the external unit-cells of a single metasurface sector require such a large phase-shift value. Thus, acceptable overall performance is expected.

A sketch of the final cylindrical *HMS* coating surrounding the dipole antenna is reported in Fig. 12. As can be appreciated, all cells are characterized by the same physical layout since the different phase-insertion values are achieved by properly tuning the varactors. In the zoom showing two consecutive unit-cells distributed along the *z*-axis, the biasing lines of the diodes are also reported. The network is embedded within the dielectric substrate and consists of two grounded horizontal strips (used as voltage reference), and another set of three strips connected to the top capacitive layer through a metallic via for setting the polarization voltages. The top strips are bridged to the $V_n$ (with $n = 1,2,3$) through a single via that is alternatively connected to the polarizing strips depending on the desired phase-insertion introduced by the cell. It is worth noticing that the ultra-thin biasing lines are cross-polarized *w.r.t.* to the antenna. Hence, the coupling with the field radiated by the source and the inductive/capacitive strips is almost negligible.

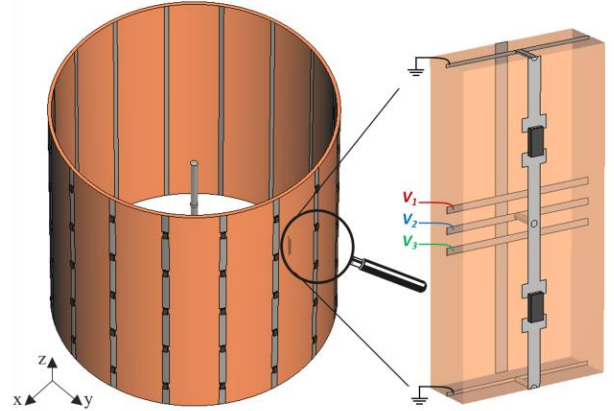

Fig. 12: Sketch of the realistic *HMS* coating surrounding the dipole antenna introducing a phase-gradient on the *xoy* plane. The *HMS* consists of a dual-layer metasurface made of inductive metallic strips and capacitive strips loaded with varactor diodes. In the inset, the zoom of two vertically consecutive unit-cells showing the biasing network layout are reported.

In Fig. 13, the numerical results of the coated antenna evaluated through full-wave simulations are reported. The gain polar plot shows that by adjusting the biasing voltages for setting the $C_{j,n}$ in the range [0.05 – 0.13 pF], four main radiating beams appear, with a maximum gain of 6.34 dBi. Remarkably, when setting the junction capacitances of all the varactor to 0.085 pF, the original omnidirectional pattern of the antenna is restored, emphasizing the reconfiguration capability of the system. Furthermore, the possibility to control the phase-insertion of the single cells allow to dynamically rotate the radiation pattern on the antenna *H*-plane, further expanding the scanning and steering capability of the device. At the same time, on the *E*-plane the shape of the radiation pattern is unaltered compared to the bare antenna scenario, as can be appreciated from the 3D radiation pattern in Fig. 14.

Albeit the presence of the *HMS* coating around the linear antenna, the matching performances are quite good, as shown in Fig. 15, where the reflection coefficient magnitude at the antenna input port is reported and compared with the one of the bare dipole. A slight shift of the central operation frequency and a reduction of the bandwidth can be observed, although at $f_0$ still good matching is achieved. The performance deterioration is mostly due to the non-ideal transmission performance of the metasurface unit-cells that partially reflect the field radiated by the source, inducing secondary currents that modify the original behavior.

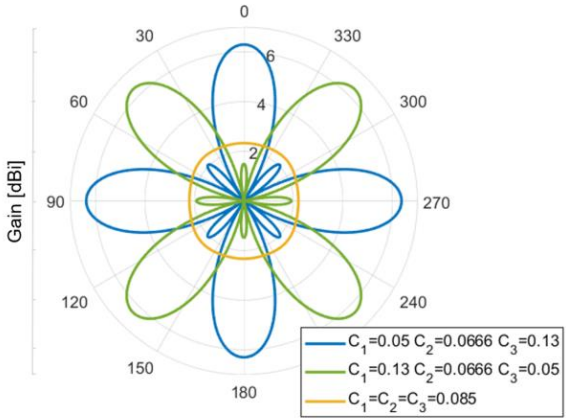

Fig. 13: Gain polar plots evaluated on the *H*-plane of the antenna for different combinations of the varactors' junction capacitance $C_{j,n}$. The values of the junction capacitances are expressed in [pF].

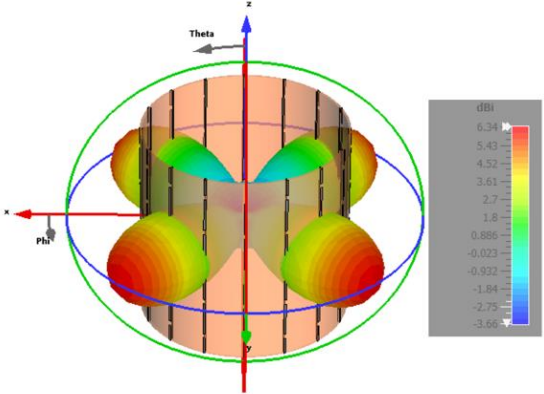

Fig. 14: 3D gain radiation pattern of the antenna for the varactors' junction capacitances combination ensuring the four-beam configuration.

Finally, in Fig. 16, the performance of the realistic *HMS* configuration is compared to the ideal setup discussed in Section III. As expected, a deterioration of the gain response of the realistic design can be noticed, with an almost 1 dB reduction of the maximum gain. However, this reduction was expected and reasonable considering the simplicity of the ideal layout, the inclusion of the biasing network, and the presence of losses coming from both the dielectric and the varactors. Moreover, we underline that, in the above full-wave simulations, the internal resistance of the varactors was included ($R_s = 2.65\,\Omega$ from the varactor datasheet), although not considered in a first instance in the circuit schematic in Fig. 11. Thanks to the reduced number of varactors, the low value of $R_s$, and the non-resonant behavior of the impedance sheets, the radiation efficiency of the antenna is quite high and equal to $e_{rad} = 0.9$, confirming the benefit of the proposed design strategy.

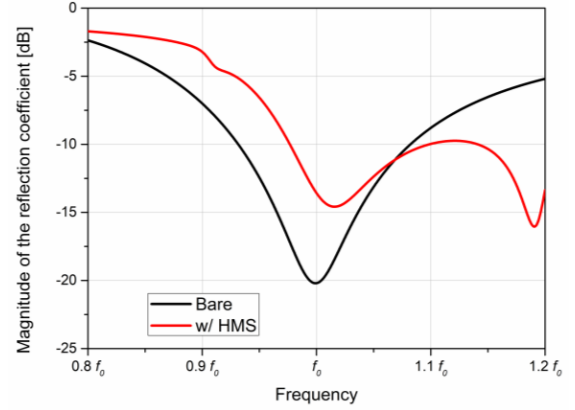

Fig. 15: Magnitude of the reflection coefficient at the antenna input port for the scenarios w/o (*black line*) and w/ (*red line*) the *HMS* coating once the varactors are set to ensure the four-beam configuration.

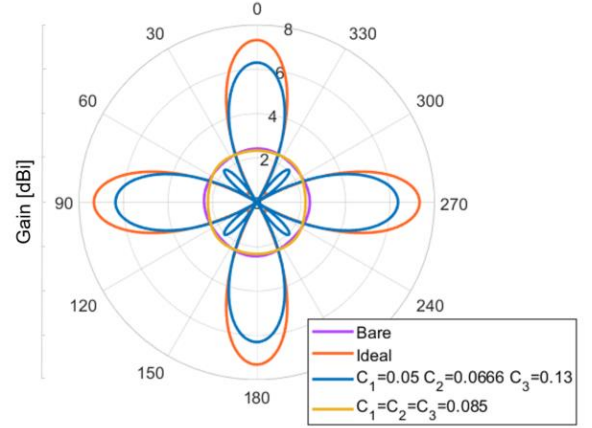

Fig. 16: Gain polar plots evaluated on the *H*-plane of the antenna when the junction capacitances of the varactors are set to achieve a four-beam pattern (*blue line*) or an omnidirectional pattern (*orange line*). For comparison, the equivalent configuration when considering ideal *HMS* unit-cell is reported (*red line*), as well as the original dipole antenna pattern (*purple line*).

## V. Conclusion

In this paper, we have shown that beamforming functionality can be introduced in linear antenna systems by means of phase-gradient *HMS* surrounding a single simple radiator like a dipole. In particular, we have shown that the original omnidirectional radiation pattern of the dipole antenna can be turned into multiple- or single-beam shapes through a judicious design of the *HMS* and, thus, by properly engineering its phase-insertion profile.

At the scope, a complete design procedure has been developed, highlighting the working principle of the structure and deriving closed-form formulas for identifying the required phase-insertion profile. In addition, a semi-analytical procedure for designing the *HMS* characterized by the desired phase-gradient profile has been reported, exploiting a conventional network approach. The effectiveness of the proposed design workflow has been assessed by using it for the synthesis of cylindrical *HMSs* enabling different beamforming effects. Finally, a realistic configuration of a phase-gradient *HMS* coating for dynamic beamforming functionalities has been presented. The reconfigurability of the system was achieved by loading the *HMS* with varactor diodes able to tune the phase-insertion exhibited by the individual unit-cells. For the unit-cell design, particular emphasis has been put on the minimization of the losses coming from the use of commercial electronic elements. This approach allowed achieving good efficiency and reasonable overall performance, as confirmed by the final numerical simulation results.

The possibility to dynamically shape the radiation pattern of such a simple antenna could find interesting applications in both outdoor and indoor wireless communications. Indeed, the device could be used in smart radio base-stations and/or smart repeaters for dynamically re-routing the signal towards specific directions or multiple users, according to the operative and environmental conditions. Moreover, these findings confirm the effectiveness of conformal cylindrical metasurfaces as a promising solution for implementing "*intelligent*" antennas capable of reconfiguring their frequency, scattering, and radiation responses.

REFERENCES

[1] L. Bariah et al., "A Prospective Look: Key Enabling Technologies, Applications and Open Research Topics in 6G Networks," *IEEE Access*, vol. 8, pp. 174792-174820, 2020.
[2] J. Zhang, E. Björnson, M. Matthaiou, D. W. K. Ng, H. Yang and D. J. Love, "Prospective Multiple Antenna Technologies for Beyond 5G," *IEEE J. Sel. Areas Commun.*, vol. 38, no. 8, pp. 1637-1660, Aug. 2020.
[3] A. Ijaz et al., "Enabling Massive IoT in 5G and Beyond Systems: PHY Radio Frame Design Considerations," *IEEE Access*, vol. 4, pp. 3322-3339, 2016.
[4] H. Frank, C. Colman-Meixner, K. D. R. Assis, S. Yan and D. Simeonidou, "Techno-Economic Analysis of 5G Non-Public Network Architectures," *IEEE Access*, vol. 10, pp. 70204-70218, 2022.
[5] T. S. Rappaport et al., "Millimeter Wave Mobile Communications for 5G Cellular: It Will Work!," *IEEE Access*, vol. 1, pp. 335-349, 2013.
[6] K. Aldubaikhy, W. Wu, N. Zhang, N. Cheng, and X. Shen, "mmWave IEEE 802.11ay for 5G fixed wireless access," *IEEE Wireless Commun.*, vol. 27, no. 2, pp. 88-95, Apr. 2020.
[7] I. K. Jain, R. Kumar and S. S. Panwar, "The impact of mobile blockers on millimeter wave cellular systems", *IEEE J. Sel. Areas Commun.*, vol. 37, no. 4, pp. 854-868, Apr. 2019.
[8] J. Du, D. Chizhik, R. Feick, M. Rodriguez, G. Castro, and R. A. Valenzuela, "Suburban fixed wireless access channel measurements and models at 28 GHz for 90% outdoor coverage," *IEEE Trans. Antennas Propag.*, vol. 68, no. 1, pp. 411-420, Jan. 2020.
[9] L. Chiaraviglio et al., "Planning 5G Networks Under EMF Constraints: State of the Art and Vision," *IEEE Access*, vol. 6, pp. 51021-51037, 2018.
[10] L. Chiaraviglio, C. Di Paolo and N. Blefari-Melazzi, "5G Network Planning Under Service and EMF Constraints: Formulation and Solutions," *IEEE Trans Mob Comput*, vol. 21, no. 9, pp. 3053-3070, 1 Sept. 2022.
[11] M. Di Renzo et al., "Smart radio environments empowered by reconfigurable AI meta-surfaces: An idea whose time has come," *EURASIP J. Wireless Commun. Netw.*, vol. 2019, no. 1, pp. 1–20, Dec. 2019.
[12] M. Di Renzo et al., "Smart Radio Environments Empowered by Reconfigurable Intelligent Surfaces: How It Works, State of Research, and The Road Ahead," *IEEE J. Selected Areas Comm.*, vol. 38, no. 11, pp. 2450-2525, Nov. 2020.
[13] F. Yang, D. Erricolo, A. Massa, "Special Issue on Smart Electromagnetic Environment," *IEEE Trans. Antennas Propag.*, vol. 69, no. 3, pp. 1838-1838, 2021.
[14] A. Massa, A. Benoni, P. Da Ru, S. K. Goudos, B. Li, G. Oliveri, A. Polo, P. Rocca, and M. Salucci, "Designing smart electromagnetic environments for next-generation wireless communications," *Telecom*, vol. 2, no. 2, pp. 213-221, 2021.
[15] G. Gradoni et al., ''Smart radio environments,'' 2021, arXiv:2111.08676.
[16] M. Barbuto et al., "Metasurfaces 3.0: a New Paradigm for Enabling Smart Electromagnetic Environments," *IEEE Trans. Antennas Propag.*, vol. 70, no. 10, pp. 8883-8897, Oct. 2022.
[17] E. Martini and S. Maci, "Theory, Analysis, and Design of Metasurfaces for Smart Radio Environments," *Proc. IEEE*, vol. 110, no. 9, pp. 1227-1243, Sept. 2022.
[18] A. Díaz-Rubio, S. Kosulnikov and S. A. Tretyakov, "On the Integration of Reconfigurable Intelligent Surfaces in Real-World Environments: A Convenient Approach for Estimation Reflection and Transmission," *IEEE Antennas Propag Mag.*, vol. 64, no. 4, pp. 85-95, Aug. 2022.
[19] R. Flamini et al., "Toward a Heterogeneous Smart Electromagnetic Environment for Millimeter-Wave Communications: An Industrial Viewpoint," *IEEE Trans. Antennas Propag.*, vol. 70, no. 10, pp. 8898-8910, Oct. 2022.
[20] G. Oliveri, P. Rocca, M. Salucci and A. Massa, "Holographic Smart EM Skins for Advanced Beam Power Shaping in Next Generation Wireless Environments," *IEEE J. Multiscale Multiphysics*, vol. 6, pp. 171-182, 2021.
[21] G. Oliveri, et al. "Building a smart EM environment - AI-Enhanced aperiodic micro-scale design of passive EM skins," *IEEE Trans. Antennas Propag.*, 2022, (early access) doi: 10.1109/TAP.2022.3151354
[22] E. Basar, M. Di Renzo, J. De Rosny, M. Debbah, M. -S. Alouini and R. Zhang, "Wireless Communications Through Reconfigurable Intelligent Surfaces," *IEEE Access*, vol. 7, pp. 116753-116773, 2019.
[23] M. Di Renzo, et al., "Reconfigurable intelligent surfaces vs. relaying: Differences, similarities, and performance comparison," IEEE Open J. Comm. Soc., vol. 1, pp. 798-807, 2020.
[24] M. A. ElMossallamy, H. Zhang, L. Song, K. G. Seddik, Z. Han and G. Y. Li, "Reconfigurable Intelligent Surfaces for Wireless Communications: Principles, Challenges, and Opportunities," *IEEE Trans. Cogn. Commun. Netw.*, vol. 6, no. 3, pp. 990-1002, Sept. 2020.
[25] V. Degli-Esposti, E. M. Vitucci, M. Di Renzo and S. Tretyakov, "Reradiation and Scattering from a Reconfigurable Intelligent Surface: A General Macroscopic Model," *IEEE Trans. Antennas Propag.*, 2022 (early access), doi: 10.1109/TAP.2022.3149660.
[26] M. Di Renzo and S. Tretyakov, "Reconfigurable Intelligent Surfaces," *Proc. IEEE*, vol. 110, no. 9, pp. 1159-1163, Sept. 2022.
[27] R. Liu, Q. Wu, M. Di Renzo and Y. Yuan, "A Path to Smart Radio Environments: An Industrial Viewpoint on Reconfigurable Intelligent Surfaces," *IEEE Wirel Commun.*, vol. 29, no. 1, pp. 202-208, Feb. 2022.
[28] S. Vellucci, A. Monti, M. Barbuto, A. Toscano, F. Bilotti, "Progress and perspective on advanced cloaking metasurfaces: from invisibility to intelligent antennas," *EPJ Appl. Metamaterials*, vol. 8, pp. 7, 2021.
[29] M. Barbuto, et al. "Intelligence enabled by 2D metastructures in antennas and wireless propagation systems," *IEEE Open J. Antennas Propag.*, vol. 3, pp. 135-153, 2022.
[30] A. Monti et al., "Overcoming mutual blockage between neighboring dipole antennas using a low-profile patterned metasurface," *IEEE Antennas Wirel. Propag. Lett.*, vol. 11, pp. 1414-1417, 2012.
[31] Z. H. Jiang, P. E. Sieber, L. Kang, and D. H. Werner, "Restoring Intrinsic Properties of Electromagnetic Radiators Using Ultralightweight Integrated Metasurface Cloaks," *Adv. Funct. Mater.*, vol. 25, pp. 4708–4716, 2015.
[32] A. Monti, J. Soric, M. Barbuto, D. Ramaccia, S. Vellucci, F. Trotta, A. Alù, A. Toscano, and F. Bilotti, "Mantle cloaking for co-site radio-frequency antennas," *Appl. Phys. Lett.*, vol. 108, 113502, 2016.
[33] S. Vellucci, A. Monti, M. Barbuto, A. Toscano, and F. Bilotti, "Satellite Applications of Electromagnetic Cloaking," *IEEE Trans. Antennas Propag.*, vol. 65, pp. 4931–4934, 2017.
[34] H. Mehrpour Bernety, A. B. Yakovlev, H. G. Skinner, S. -Y. Suh and A. Alù, "Decoupling and Cloaking of Interleaved Phased Antenna Arrays Using Elliptical Metasurfaces," *IEEE Trans. Antennas Propag*, vol. 68, no. 6, pp. 4997-5002, June 2020.

[35] Z. H. Jiang, M. D. Gregory and D. H. Werner, "A Broadband Monopole Antenna Enabled by an Ultrathin Anisotropic Metamaterial Coating," in *IEEE Antennas Wirel. Propag. Lett.*, vol. 10, pp. 1543-1546, 2011.

[36] S. Vellucci, A. Toscano, F. Bilotti, A. Monti and M. Barbuto, "Coating Metasurfaces Enabling Antenna Frequency Reconfigurability for Cognitive Radio System," *2021 IEEE International Symposium on Antennas and Propagation and USNC-URSI Radio Science Meeting (APS/URSI)*, pp. 417-418, 2021.

[37] S. Vellucci, et al. "Multi-layered coating metasurfaces enabling frequency reconfigurability in wire antenna," *IEEE Open J. Antennas Propag.*, vol. 3, pp. 206-216, 2022.

[38] A. Monti, M. Barbuto, A. Toscano, and F. Bilotti, "Nonlinear Mantle Cloaking Devices for Power-Dependent Antenna Arrays," *IEEE Antennas Wirel. Propag. Lett.*, vol. 16, pp. 1727-1730, 2017.

[39] S. Vellucci et al., "Non-linear Mantle Cloaks for Self-Configurable Power-Dependent Phased Arrays," *2020 XXXIIIrd General Assembly and Scientific Symposium of the International Union of Radio Science*, pp. 1-3, 2020.

[40] S. Vellucci et al., "On the Use of Non-Linear Metasurfaces for Circumventing Fundamental Limits of Mantle Cloaking for Antennas," *IEEE Trans. Antennas Propag.*, vol. 69, no. 8, pp. 5048-5053, 2021.

[41] S. Vellucci, A. Monti, M. Barbuto, A. Toscano, F. Bilotti "Waveform-Selective Mantle Cloaks for Intelligent Antennas," *IEEE Trans. Antennas Propag.*, vol. 68, no. 3, pp. 1717-1725, Mar. 2020.

[42] D. Ushikoshi, "Pulse-driven self-reconfigurable meta-antennas," *Nat. Commun.*, vol. 14, no. 633, 2023.

[43] C. Balanis, *Antenna Theory: Analysis and Design*, John Wiley & Sons, Hoboken, NJ, USA, 2005.

[44] C. Pfeiffer and A. Grbic, "Metamaterial Huygens' surfaces: Tailoring wave fronts with reflectionless sheets," *Phys. Rev. Lett.*, vol. 110, p. 197401, 2013.

[45] M. Selvanayagam and G. V. Eleftheriades, "Discontinuous electromagnetic fields using orthogonal electric and magnetic currents for wavefront manipulation," *Opt. Express*, vol. 21, no. 12, pp. 14 409–14 429, 2013.

[46] A. Monti, A. Alù, A. Toscano, and F. Bilotti, "Surface impedance modeling of all-dielectric metasurfaces," *IEEE Trans. Antennas Propag.*, vol. 68, No. 3, pp. 1799-1811, 2019.

[47] V. S. Asadchy, M. Albooyeh, S. N. Tcvetkova, A. Díaz-Rubio, Y. Ra'di, and S. A. Tretyakov, "Perfect control of reflection and refraction using spatially dispersive metasurfaces," *Phys. Rev. B*, vol. 94, no. 7, 2016.

[48] C. Huygens, *Traité de la Lumière*, Pieter van der Aa, Leyden, 1690.

[49] M. Chen, E. Abdo-Sánchez, A. Epstein, and G. V. Eleftheriades, "Theory, design, and experimental verification of a reflectionless bianisotropic Huygens' metasurface for wide-angle refraction," *Phys. Rev. B*, 97, 125433, 2018.

[50] A. Epstein and G. V. Eleftheriades, "Huygens' metasurfaces via the equivalence principle: design and applications," *J. Opt. Soc. Am. B*, vol. 33, A31-A50, 2016.

[51] G. A. Egorov and G. V. Eleftheriades, "Theory and Simulation of Metasurface Lenses for Extending the Angular Scan Range of Phased Arrays," *IEEE Trans. Antennas Propag.*, vol. 68, no. 5, pp. 3705-3717, May 2020.

[52] Y. -H. Lv, X. Ding, B. -Z. Wang and D. E. Anagnostou, "Scanning Range Expansion of Planar Phased Arrays Using Metasurfaces," *IEEE Trans. Antennas Propag.*, vol. 68, no. 3, pp. 1402-1410, March 2020.

[53] A. Monti et al., "Quadratic-Gradient Metasurface-Dome for Wide-Angle Beam-Steering Phased Array With Reduced Gain Loss at Broadside *IEEE Trans. Antennas Propag.*, vol. 71, no. 2, pp. 2022-2027, Feb. 2023.

[54] V. G. Ataloglou et al., "Static and Reconfigurable Huygens' Metasurfaces: Use in Antenna Beamforming and Beam Steering," *IEEE Antennas Propag. Mag.*, vol. 64, no. 4, pp. 73-84, 2022.

[55] D. Ramaccia et al., "Metasurface Dome for Above-the-Horizon Grating Lobes Reduction in 5G-NR Systems," *IEEE Antennas Wirel. Propag. Lett.*, vol. 21, no. 11, pp. 2176-2180, 2022.

[56] S. Vellucci, M. Longhi, A. Monti, M. Barbuto, A. Toscano and F. Bilotti, "Antenna Pattern Shaping through Functionalized Metasurface Coatings," *2022 Sixteenth International Congress on Artificial Materials for Novel Wave Phenomena (Metamaterials)*, pp. 466-468, 2022.

[57] C. Pfeiffer et al., "Efficient Light Bending with Isotropic Metamaterial Huygens' Surfaces," *Nano Lett.*, vol. 14, pp. 2491–2497, 2014.

[58] A. Epstein and G. V. Eleftheriades, "Arbitrary power-conserving field transformations with passive lossless omega-type bianisotropic metasurfaces," *IEEE Trans. Antennas Propag.*, vol. 64, pp. 3880–3895, 2016.

[59] D.M. Pozar, *Microwave Engineering.* John Wiley & Sons Inc, 2011.

[60] A. Monti et al., "Optimal Design of Huygens Metasurfaces for Oblique Incidence through a Microwave Network Approach," *2022 Microwave Mediterranean Symposium (MMS)*, pp. 1-4, 2022.

[61] D. A. Frickey, "Conversions between S, Z, Y, H, ABCD, and T parameters which are valid for complex source and load impedances," *IEEE Trans. Microw. Theory Tech.*, vol. 42, no. 2, pp. 205-211, 1994.

[62] G. Bertin, F. Bilotti, B. Piovano, R. Vallauri, L. Vegni, "Switched beam antenna employing metamaterial-inspired radiators," *IEEE Trans. Antennas Propag.*, vol. 60, pp. 3583-3593, 2012.

[63] R. Harrington, "Reactively controlled directive arrays," *IEEE Trans. Antennas Propag.*, vol. 26, no. 3, pp. 390-395, May 1978.

[64] S. Tretyakov, *Analytical Modeling in Applied Electromagnetics*. Norwood, MA, USA: Artech House, 2003.

[65] A. Monti, J.C. Soric, A. Alù, A. Toscano, and F. Bilotti, "Anisotropic Mantle Cloaks for TM and TE Scattering Reduction," *IEEE Trans. Antennas Propag.*, vol. 63, no. 4, pp. 1775-1788, 2015.

[66] C. Qi and A. M. H. Wong, "Discrete Huygens' Metasurface: Realizing Anomalous Refraction and Diffraction Mode Circulation With a Robust, Broadband and Simple Design," *IEEE Trans. Antennas Propag.*, vol. 70, no. 8, pp. 7300-7305, 2022.